\documentclass[twocolumn,prl,amsmath,letterpaper,floatfix,showpacs]{revtex4}
\pdfoutput=1 
\usepackage{graphicx}
\usepackage{amsmath}
\usepackage{amssymb}
\usepackage{color}

\newcommand{\figref}[2]{Fig.~\ref{#1}~{\bf #2}}
\newcommand{\figrefs}[1]{Fig.~\ref{#1}}

\begin{document}

\title{Observation of nondispersing classical-like molecular rotation}

\author{Aleksey Korobenko$^{1}$, John W. Hepburn$^{1,2}$, Valery Milner$^{1}$}
\affiliation{$^{1}$Department of  Physics \& Astronomy, The University of British Columbia, Vancouver, Canada}
\affiliation{$^{2}$Department of  Chemistry, The University of British Columbia, Vancouver, Canada}

\begin{abstract}
Using the technique of an optical centrifuge, we produce rotational wave packets which evolve in time along either classical-like or non-classical trajectories. After releasing O$_{2}$ and D$_{2}$ molecules from the centrifuge, we track their field-free rotation by monitoring the molecular angular distribution with velocity map imaging. Due to the dispersion of the created rotational wave packets in oxygen, we observe a gradual transition between ``dumbbell''-shaped and ``cross''-shaped distributions, both rotating with a classical rotation frequency. In deuterium, a much narrower rotational wave packet is produced and shown to evolve in a truly classical non-dispersing fashion.
\end{abstract}
\maketitle

\section{Introduction}
Microscopic quantum objects behaving in a classical manner are of great interest due to the fundamental aspects of quantum-classical correspondence. The well known examples include coherent states of the quantum harmonic oscillator, slowly spreading Rydberg wave packets\cite{Yeazell88} and nonspreading Trojan wave packets\cite{Maeda05} in highly excited atoms. Classically behaving quantum wave packets preserve their minimum uncertainty shape as they move along the corresponding classical trajectories\cite{Parker86, Buchleitner1995, Kalinski1996}. In the case of rotating molecules, classical-like behavior implies nondispersing rotation of the molecular wave function with narrow distribution of angular momenta and strong localization around the internuclear axis.

Angular localization of molecular axes, well known as molecular alignment\cite{Stapelfeldt03}, has been long recognized as an important tool for controlling chemical reactions in gaseous media\cite{Zare1998} and at gas-surface interfaces\cite{Kuipers1988, Shreenivas2010}, for imaging molecular orbitals\cite{Itatani2004} and generating extreme UV radiation\cite{Itatani2005, Wagner2007}, for deflecting molecular beams\cite{Purcell2009} and separating molecular isotopes\cite{Fleischer06}. Although the popular non-adiabatic approach to molecular alignment with intense non-resonant laser pulses can produce very narrow angular distributions\cite{Seideman1995, Rosca02, Dooley03, Leibscher2003, Renard04, Underwood05, Cryan09, Zhdanovich12}, the angular momentum of the aligned molecules is usually poorly defined, both in magnitude and direction. Hence, the created rotational wave packet quickly disperses in angle, undergoing oscillations between aligned and non-aligned distributions. Controlling the direction of molecular rotation with a sequence of two\cite{Fleischer09, Kitano09} or more\cite{Zhdanovich11,Bloomquist12} laser pulses offers better defined angular momentum, but suffers from the lack of selectivity and efficiency.
\begin{figure}[b]
\includegraphics[width=\columnwidth]{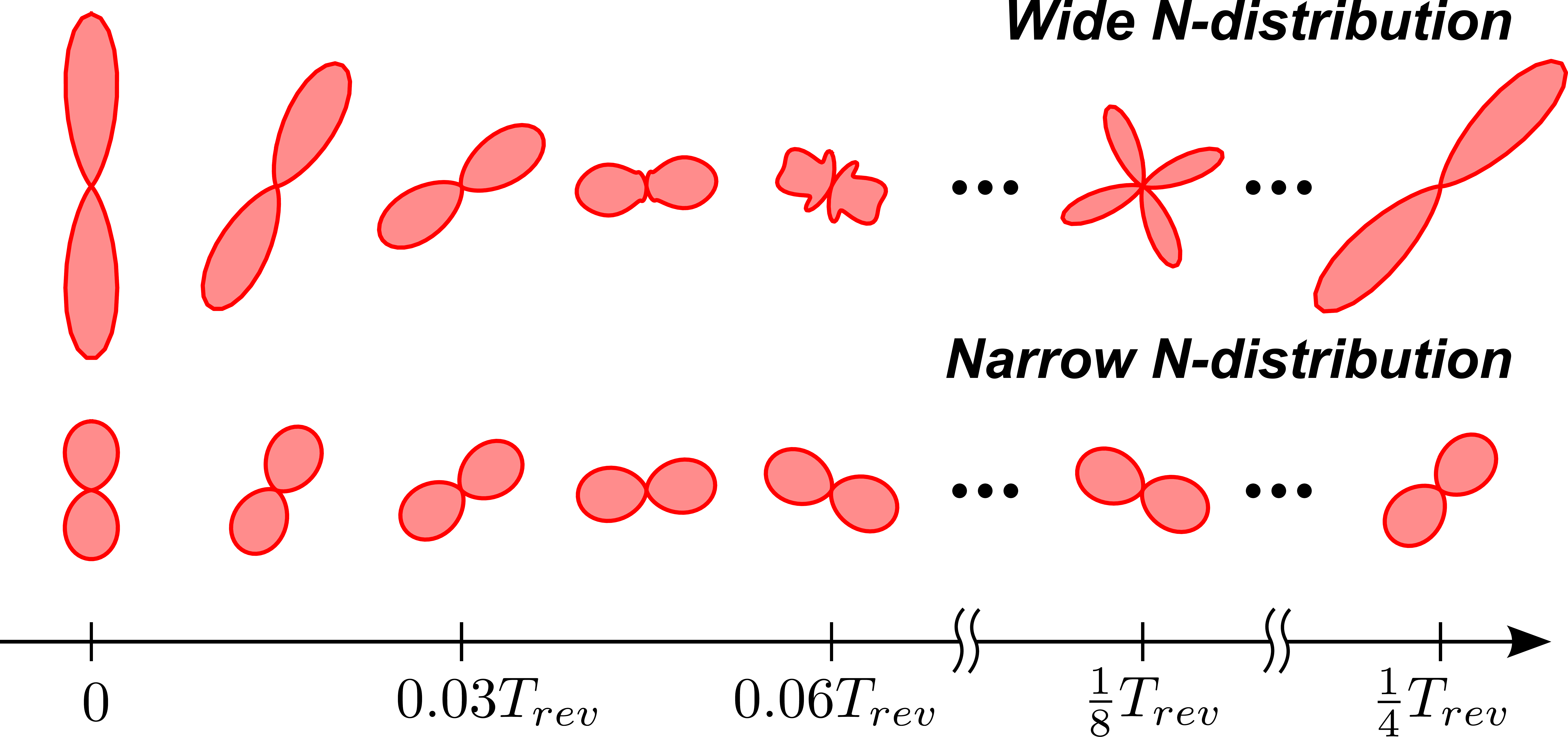}
\caption{Time evolution of a wide (top) and a narrow (bottom) rotational wave packet, consisting of many or only two rotational states, respectively. The latter is an example of a quantum ``cogwheel state'' described in text. $T_{rev}$ is the rotational revival time. }
\label{fig_time_evol}
\end{figure}

In this work, we use an optical centrifuge as a much more selective and efficient method of creating localized rotational wave packets\cite{Karczmarek1999, Villeneuve00, Yuan11, Korobenko2014a}. The centrifuge technique enables us to execute a complete transfer of molecules from the initial ground rotational level to a coherent superposition of rotational states $| N,M_N=N \rangle$, where $N$ (either only even or only odd) and $M_{N}$ are the rotational and magnetic quantum numbers, respectively. If this wave packet is initially localized along a certain spatial axis (upper left picture in Fig.\ref{fig_time_evol}), it will rotate with an angular frequency of $\Omega_N=(E_{N+2}-E_N)/(2\hbar)$ (where $E_{N}$ is the rotational energy) while simultaneously undergoing angular dispersion. For a homonuclear molecule with the rotational constant $B$, it will then revive to its initial shape, up to a rotation by a constant angle, at one quarter of the revival time $T_{rev}=h/(2B)$ (upper right picture in Fig.\ref{fig_time_evol}). The so-called fractional revivals occur at intermediate times $t=\frac{p}{q}T_{rev}$, with $\frac{p}{q}$ being an irreducible fraction\cite{Averbukh1989, Averbukh1991}. Provided the number of populated rotational states is large, at the time of a fractional revival the initial wave packet develops $\frac{q}{4}(3-(-1)^q)$ uniformly distributed lobes, which keep rotating with the same frequency $\Omega_N$. For example, around $t=T_{rev}/8$, the angular distribution exhibits four peaks in the plane of rotation, as shown in the top row of \figrefs{fig_time_evol}. As a \textit{transient} effect, such multi-axial alignment has been observed experimentally\cite{Dooley03} and used as a main ingredient in the new scheme for quantum logic gates based on the control of rotational wave packets\cite{Lee2004, Shapiro2005}.
\begin{figure*}[ht!]
\includegraphics[width=2\columnwidth]{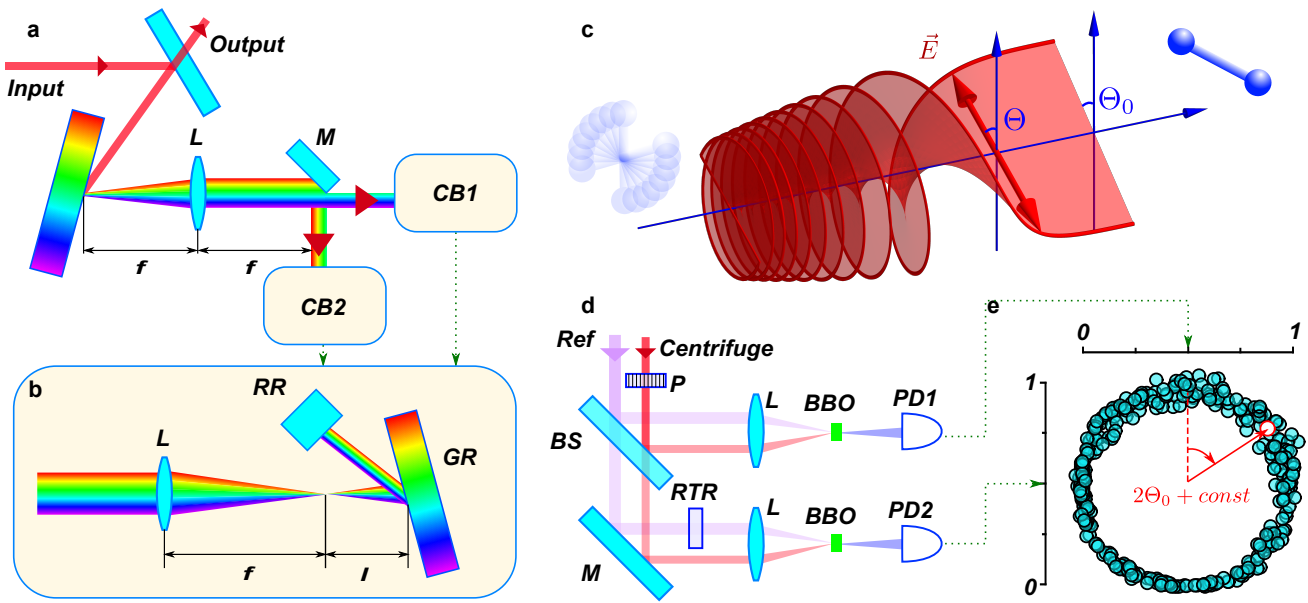}
\caption{\textbf{(a)} Optical centrifuge shaper, consisting of the following elements: diffraction gratings (GR), lenses (L), mirrors (M), retro-reflectors (RR) and two ``chirp boxes'' (CB1 and CB2), shown in panel \textbf{(b)}. \textbf{(c)}Illustration of the electric field of an optical centrifuge. Linear polarization of the field follows a ``corkscrew'' surface, adiabatically exciting molecules to high rotational states. The shape of the centrifuge is constant, while its orientation in the laboratory frame, $\Theta_0$, is changing randomly from pulse to pulse. \textbf{(d)} Optical setup for measuring the orientation angle $\Theta_0$. Labels indicate the following components: ultrashort reference pulse (Ref), beam splitter (BS), polarizer (P), lenses (L), photodiodes (PD1 and PD2), BaB$_2$O$_4$ crystals (BBO) and glass plate retarder (RTR). \textbf{(e)} Experimentally detected correlation between the two photodiode signals. Each point corresponds to a different orientation angle of the centrifuge, as indicated.}
\label{fig_centrifuge}
\end{figure*}
If the wave packet consists of only two states $|N,M_{N}=N\rangle$ and $|N+2n,M_{N}=N+2n\rangle$, where $n$ is an integer, the resulting ``cogwheel state'' rotates without spreading, preserving its shape \textit{at all times}\cite{Lapert2011}. The wave function of a quantum cogwheel is symmetric with respect to rotation by $\pi /n$ and follows the classical-like nondispersing motion indefinitely\cite{Yun2013}. The simplest case of a cogwheel state with $n=1$ is shown in the lower row of \figrefs{fig_time_evol}. Similarly to the importance of transient cogwheel distributions in quantum information processing\cite{Shapiro2005}, it has been speculated that nondispersing cogwheel states could open new perspectives in the development of molecular machines, whereas in metrology, they could serve as ``molecular stopwatches'' on a femtosecond time scale with immediate applications in synchronizing ultrafast laser sources\cite{Cryan2011}. Here, we report on the first direct observation of the nondispersing $n=1$ cogwheel state in centrifuged D$_{2}$ molecules. In oxygen, we create a coherent superposition of more than two rotational states and demonstrate the transition between the classical-like rotation of a molecular stopwatch and that of a quantum cogwheel.

\section{Experimental setup}
To produce the electromagnetic field of an optical centrifuge, we use a pulse shaper schematically shown in \figref{fig_centrifuge}{a}. Mirror M, positioned in the Fourier plane of the input lens L, splits in half the spectrum of 800 nm, 35 fs (full width at half-maximum, FWHM) pulses from a Ti:Sapphire regenerative amplifier. Frequency chirps of equal magnitude ($\beta =$0.26 ps$^{-2}$) and opposite signs are then applied to the two spectral halves by means of the two ``chirp boxes'' (CB1 and CB2), described in \figref{fig_centrifuge}{b}. Circularly polarized with an opposite handedness, they are finally re-combined into a single beam to interfere and produce a linearly polarized pulse, whose plane of polarization undergoes an accelerated rotation along a ``corkscrew''-shaped surface illustrated in \figref{fig_centrifuge}{c}. Interacting with an induced dipole moment, the centrifuge field aligns molecular axes along the direction of its polarization and forces the molecules to spin along with it. The terminal frequency of this rotation is controlled by truncating the spectral bandwidth of the centrifuge pulses, and can reach as high as 10 THz for our laser system\cite{Korobenko2014a}.
\begin{figure*}[ht!]
\includegraphics[width=2\columnwidth]{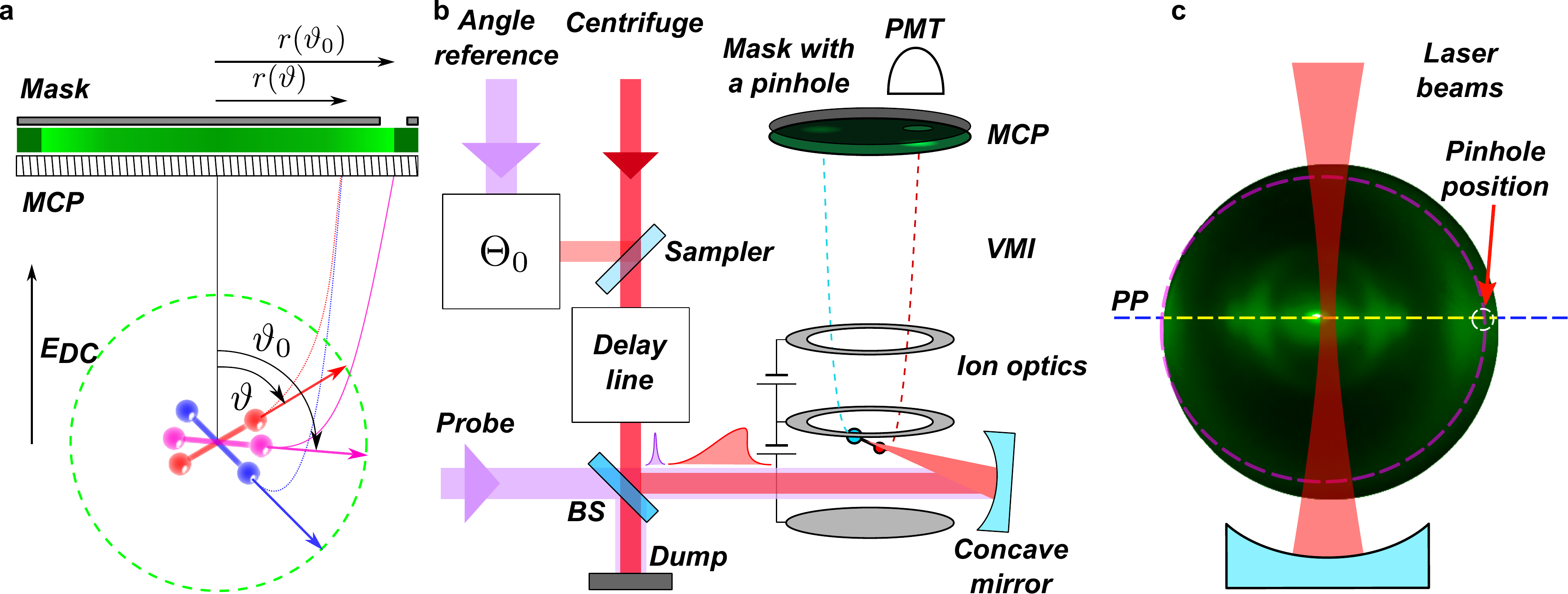}
\caption{\textbf{(a)} Illustration of the operation principle of Velocity Map Imaging (VMI) used for determining the molecular orientation. $E_\textsc{DC}$ indicates the direction of the permanent electric field used for accelerating ions toward the multi-channel plate (MCP). See text for other details. \textbf{(b)} Experimental setup. Centrifuge and probe pulses are combined on a beam splitter (BS) and sent into a vacuum chamber, where both beams are focused on a molecular jet with a concave mirror. Relative timing of the two pulses is controlled with a precision delay line. Angular orientation, $\Theta_0$, of each centrifuge pulse is determined in an optical setup described in text and \figref{fig_centrifuge}{d}. VMI setup consists of the standard time-of-flight ion optics, a multi-channel plate and a phosphorus screen. Signal from a small region of the phosphorus screen, behind a pinhole in an opaque mask, is measured with a photomultiplier tube (PMT). \textbf{(c)} Example of a VMI image. `PP' indicates the polarization plane of circularly polarized probe pulses. See text for details.}
\label{fig_setup}
\end{figure*}

Key to this work (as will become clear below) is the ability to determine the initial orientation angle $\Theta_0$ of the centrifuge (\figref{fig_centrifuge}{c}). Due to the lack of interferometric stability between the two long arms of the centrifuge shaper, this angle changes randomly from pulse to pulse. We measure $\Theta_0$ by directing a small part of the centrifuge pulse into the optical setup shown in \figref{fig_centrifuge}{d}. First, the rotating polarization vector of the centrifuge is projected onto the vertical axis by means of a linear polarizer (P). We then measure the intensity of the transmitted vertical component at a fixed delay time $\tau$ with respect to the rising edge of the centrifuge, using the cross-correlation optical gating method. The latter is executed by frequency mixing the (polarization-filtered) centrifuge pulse with a transform limited ultrashort reference pulse (Ref) on a nonlinear BaB$_2$O$_4$ (BBO) crystal. The cross-correlation signal, recorded with a photo-diode (PD1) is proportional to $\cos^{2}\Theta(\tau )  \propto 1+\cos 2\Theta(\tau )=I_c(\tau)$, where $\Theta(\tau)$ denotes the angle of the centrifuge field polarization at time $\tau$ (\figref{fig_centrifuge}{c}). Note, that since $\Theta(\tau)=\Theta_0+\beta\tau^2$, with the second term reflecting the accelerated rotation of the centrifuge, $I_c(\tau)=1+\cos(2\Theta_0+2\beta\tau^2)$.

To avoid the ambiguity in extracting the value of $\Theta_0$, another cross-correlation is performed simultaneously at a slightly different delay time $\tau +\Delta \tau$. The extra time difference, $\Delta \tau$, is introduced with a thin glass plate retarder (RTR). The intensity recorded with a second photo-diode (PD2) is proportional to $I_c(\tau+\Delta\tau)\approx 1+\cos(2\Theta_0+2\beta\tau^2+4\beta\tau\Delta\tau)$. We adjust $\tau$ so as to satisfy $4\beta\tau\Delta\tau=\frac{\pi}{2}$ and, therefore, $I_c(\tau+\Delta\tau)=1-\sin(2\Theta_0+2\beta\tau^2)$. By plotting the two photo-diode signals against one another, the orientation angle $\Theta_0$ can be easily extracted up to a constant angle $\beta\tau^2$, as shown in \figref{fig_centrifuge}{d}.

Having the ability to determine the initial angle, at which the molecules are released from the centrifuge, one can now keep track of their rotational dynamics by measuring the angular probability density of the rotational wave packet as a function of time. This measurement is often carried out by means of the velocity map imaging (VMI) technique \cite{Larsen1998}. In VMI, an intense femtosecond laser pulse dissociates the molecule via a multi-photon ionization process followed by a Coulomb explosion. The method enables one to map out the direction of the ion recoil and, therefore, the angular distribution of molecular axes at the moment of explosion\cite{Dooley03}. The recoiling ions are extracted with an electric field and projected onto a micro-channel plate detector equipped with a phosphorus screen, where they leave a spatially resolved fluorescence trace. The azimuthal angle of the molecular axis is mapped to the angular coordinate $\varphi$, while its polar angle $\vartheta$ - to the radial coordinate $r(\vartheta)$ of the ion image on the phosphorus screen (\figref{fig_setup}{a}).

The energy release in a Coulomb explosion is fully determined by the dissociation channel, i.e. by the final states of the recoiling ions. For any particular channel, all possible initial momenta reside on a sphere, with each point corresponding to a distinct angular orientation of the exploding molecule (\figref{fig_setup}{a}, green dashed circle). Note, that fast molecular rotation may add a non-negligible tangential component to the momentum of recoiling ions. However, for a synchronously rotating molecules with a well-defined angular momentum, as produced by an optical centrifuge, this will simply rotate the whole orientation sphere, preserving the one-to-one correspondence between the location on the sphere and the ion image on the phosphorus screen.

In our experiment, we combine a beam of circularly polarized 35 fs ionizing probe pulses with a beam of centrifuge pulses, and focus both laser beams on a supersonically expanded gas of either O$_{2}$ or D$_{2}$ molecules between the charged plates of the VMI setup (\figref{fig_setup}{b}). The rotational temperature in the supersonic jet is below $10$~K, which in the case of oxygen means that the majority of molecules are in the ground rotational state, $N=1$. An example of a typical VMI image of Coulomb-exploded oxygen molecules is shown in \figref{fig_setup}{c}. Multiple dissociation channels are represented by a number of concentric rings, whose center is shifted to the left from the vertical axis due to the initial molecular velocity in the direction of the supersonic jet. The rings are not uniform in the plane of the image because the ionization cross-section depends on the angle between the molecular axis and the polarization plane of a circularly polarized probe pulse (PP).

In general, a single point on the VMI image may originate from two different molecular polar angles $\vartheta$, corresponding to the positive and negative vertical components of the recoil velocity, as illustrated by the blue and red lines in \figref{fig_setup}{a}. To resolve the ambiguity, we introduce an opaque mask with a pinhole, centered on the circumference of the largest observable dissociation ring (\figref{fig_setup}{c}), which results from a single molecular orientation angle, labeled as $\vartheta_{0}$ in \figref{fig_setup}{a}.

The phosphorescence signal, recorded behind this pinhole with a photomultiplier tube (PMT), is proportional to the molecular angular distribution at this orientation $I_\textsc{vmi}(t; \Theta_0)=\left\vert\Psi_{\Theta_0}(\vartheta=\vartheta_0, \varphi=0, t)\right\vert^2$, and includes the orientation angle of the centrifuge, $\Theta_0$, as a parameter. Due to the axial symmetry of the molecular system, its wave function depends only on the relative angle $\vartheta_0-\Theta_0$, which allows us to re-write the expression for the detected signal as $I_\textsc{vmi}(t; \Theta_0)=\left\vert\Psi_0(\vartheta=\vartheta_0-\Theta_0, \varphi=0,t)\right\vert^2$. Recording the latter for multiple values of (randomly changing, but accurately determined) $\Theta_0$ enables us to extract the molecular angular distribution as a function of time since the release from the centrifuge. In contrast to the conventional VMI procedure, in which the molecular orientation is typically determined from the whole two-dimensional image, our method offers a single-shot capability which is essential for tracking molecular rotation excited by an optical centrifuge.
\begin{figure}[b]
\includegraphics[width=\columnwidth]{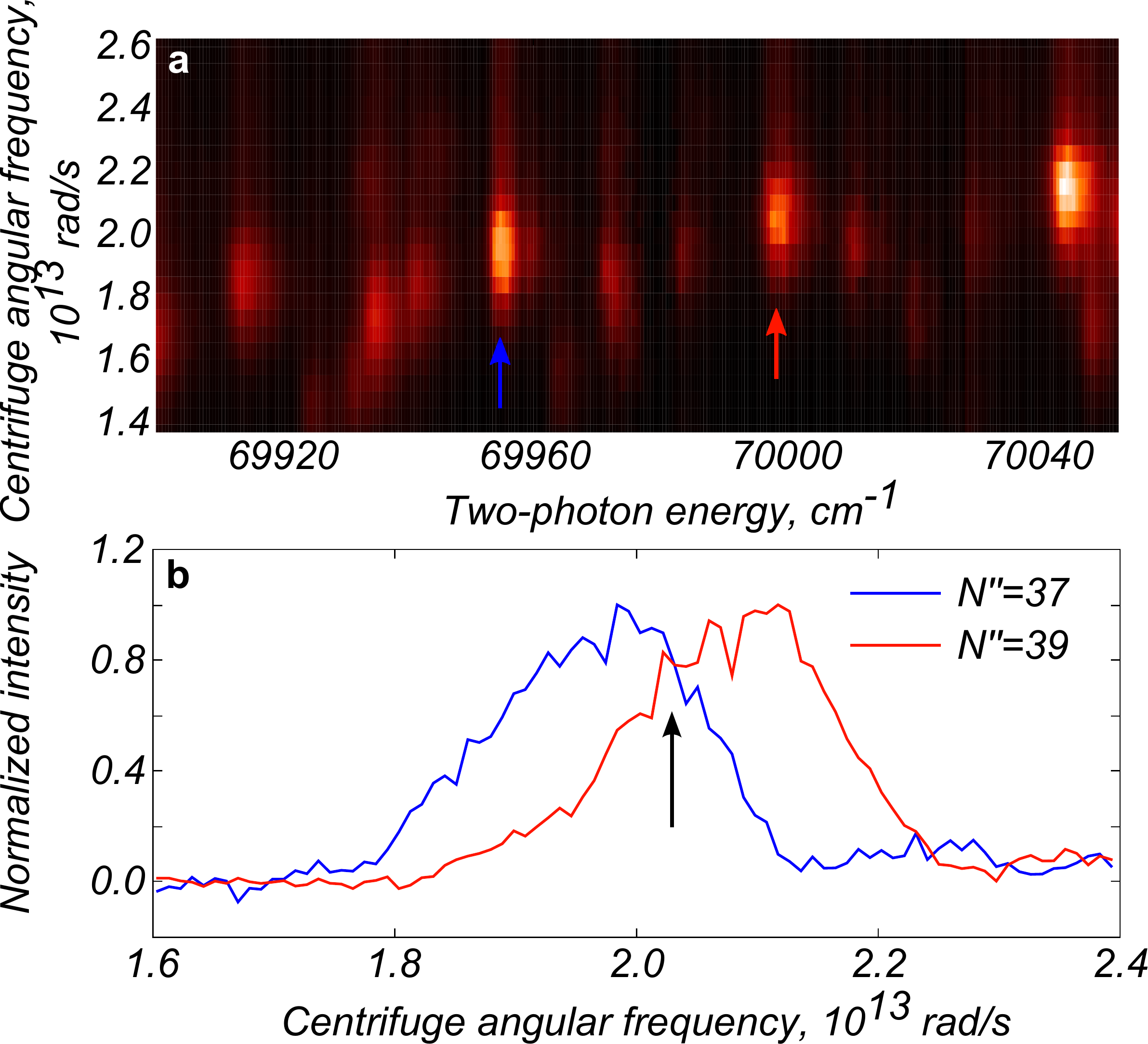}
\caption{\textbf{(a)} Two-dimensional resonance-enhanced multi-photon ionization (REMPI) spectrogram of $C^3\Pi_g(v=2)\leftarrow\leftarrow X^3\Sigma_g^-(v=0)$ two-photon transition in oxygen\cite{Korobenko2014b}. Arrows mark two resonances originated from the rotational sates with $N=37$ (blue) and $N=39$ (red). \textbf{(b)} Normalized population of the rotational states with $N=37$ (blue) and $N=39$ (red) as a function of the centrifuge final angular velocity. The black arrow shows the terminal angular frequency of the centrifuge used in this work for creating rotational wave packets in O$_{2}$.}
\label{fig_rempi}
\end{figure}

\section{Results}
To monitor the free evolution of the rotational wave packets in centrifuged oxygen, we truncate the spectral bandwidth of the centrifuge so as to match its terminal rotational frequency with that of an oxygen molecule occupying rotational states with $N=37$ and $N=39$. This is achieved by employing the technique of ``centrifuge spectroscopy'' recently developed in our group\cite{Korobenko2014b}. The method is based on resonance-enhanced multi-photon ionization (REMPI), which employs a tunable nanosecond dye laser for ionizing the centrifuged molecules via a multi-photon (2+1) resonant process. The O$_2^+$ ion yield is measured as a function of both the centrifuge rotational frequency and the two-photon energy of ns pulses.

Scanning in the appropriate range of energies, known from a thorough analysis in Ref.\citenum{Korobenko2014b}, results in a two-dimensional REMPI spectrogram shown in \figref{fig_rempi}{a}. The blue and red arrows point at the resonant transitions originating from the states with rotational quantum numbers $N=37$ and $N=39$, respectively. Vertical cross sections at the corresponding two-photon energies are plotted in \figref{fig_rempi}{b} and depict the relative populations of the two states. To make them equally populated, we fix the terminal angular frequency of the centrifuge at the value indicated by the black arrow. Note, however, that since the distance between the two peaks and their half-widths are comparable, the two neighboring states with $N=35$ and $N=41$ are also populated by the centrifuge. The population spread among multiple $N$-states stems from an imperfect adiabaticity of the centrifugal excitation\cite{Korobenko2014a}.
\begin{figure*}[t]
\includegraphics[width=2\columnwidth]{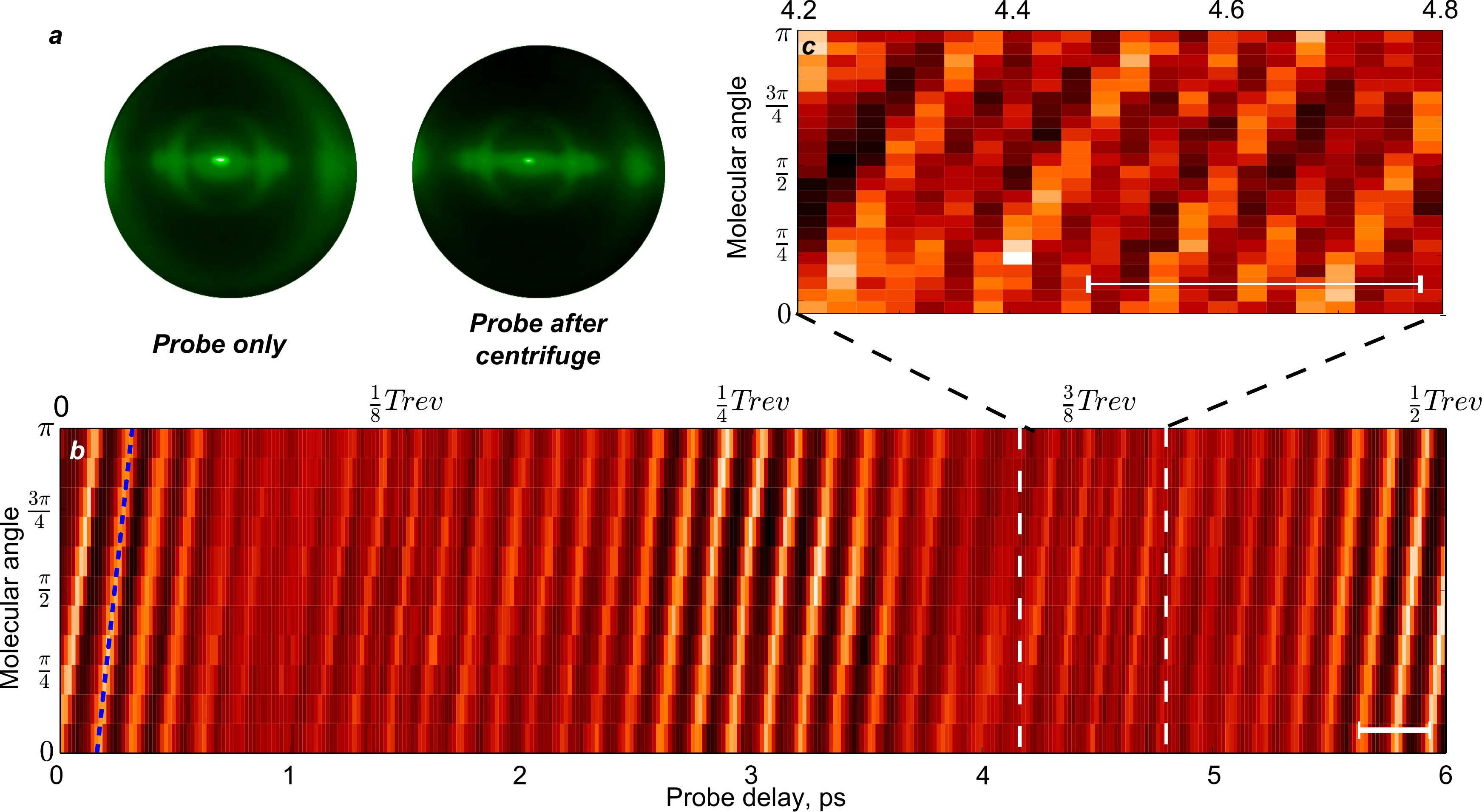}
\caption{\textbf{(a)} Ion images averaged over 10,000 shots for non-centrifuged (left) and centrifuged (right) oxygen molecules. \textbf{(b)} Probability density as a function of the molecular angle and the free propagation time. Time zero corresponds to approximately 100 ps since the release from the centrifuge. The blue dashed line marks the calculated trajectory of a ``dumbbell'' distribution rotating with the terminal angular frequency of the centrifuge, whose classical period is indicated with the white horizontal bar at the lower right corner. \textbf{(c)} Zoom-in to the region near $\frac{3}{8} T_{rev}$, taken with better averaging and angular resolution. Twice higher density of the tilted lines (4 per classical period) stems from the emergence of a ``cross''-shaped distribution with four peaks along two perpendicular directions.}
\label{fig_oxygen}
\end{figure*}

Two VMI images of oxygen, one with and one without the centrifuge, are shown in \figref{fig_oxygen}{a}. Pre-excitation with an optical centrifuge results in the visible narrowing of the ion distribution in space, owing to the localization of the rotational $\left\vert N, M_{N}=N\right\rangle$ wave functions in the plane perpendicular to the propagation direction of the centrifuge (and therefore, normal to the VMI screen). The centrifuge orientation angle $\Theta_0$, randomly changing from pulse to pulse, is measured simultaneously with the photomultiplier signal which, as described above, represents the probability density of the molecular wave function in the direction of the polar angle $\vartheta_0-\Theta _{0}$. After binning the PMT signal into 11 angle intervals, assumed by $-\Theta _0$ and uniformly distributed from 0 to $\pi $, we plot it as a function of the time delay between the centrifuge and the Coulomb-explosion pulses.

The result is shown in \figref{fig_oxygen}{b}. Being aligned at the moment of the release from the centrifuge, the molecules first undergo free rotation with the expected classical frequency of 3.3 full rotations per picosecond, indicated by the blue dashed line at the beginning of the scan. The wave packet, however, gradually disperses and, although the rotational dynamics could still be identified by the overall linear tilt of the signal, its contrast decreases dramatically. This happens because of the admixture of the neighboring $N=35, 41$ states to the rotational wave packet created around $N=37$ and $N=39$.

As expected, the alignment reappears every quarter-revival time, $\frac{1}{4} T_{rev}$ (middle section in \figref{fig_oxygen}{b}). In the vicinities of $\frac{1}{8}T_{rev}$ and $\frac{3}{8}T_{rev}$, another fractional revival is observed. After repeating the measurement around $\frac{3}{8}T_{rev}$ with better averaging and higher angular resolution, we plot the result in \figref{fig_oxygen}{c}. Here, instead of two tilted traces per one classical period (marked by a horizontal white bar), reflecting a double-peaked ``dumbbell''-shaped alignment geometry, four parallel traces per period can be seen, corresponding to the ``cross''-shaped angular distribution peaked in four spatial directions. This observation is well anticipated from the analysis of the rotational dynamics outlined in the introduction and illustrated in the top row of \figrefs{fig_time_evol}.
\begin{figure}[tb]
\includegraphics[width=\columnwidth]{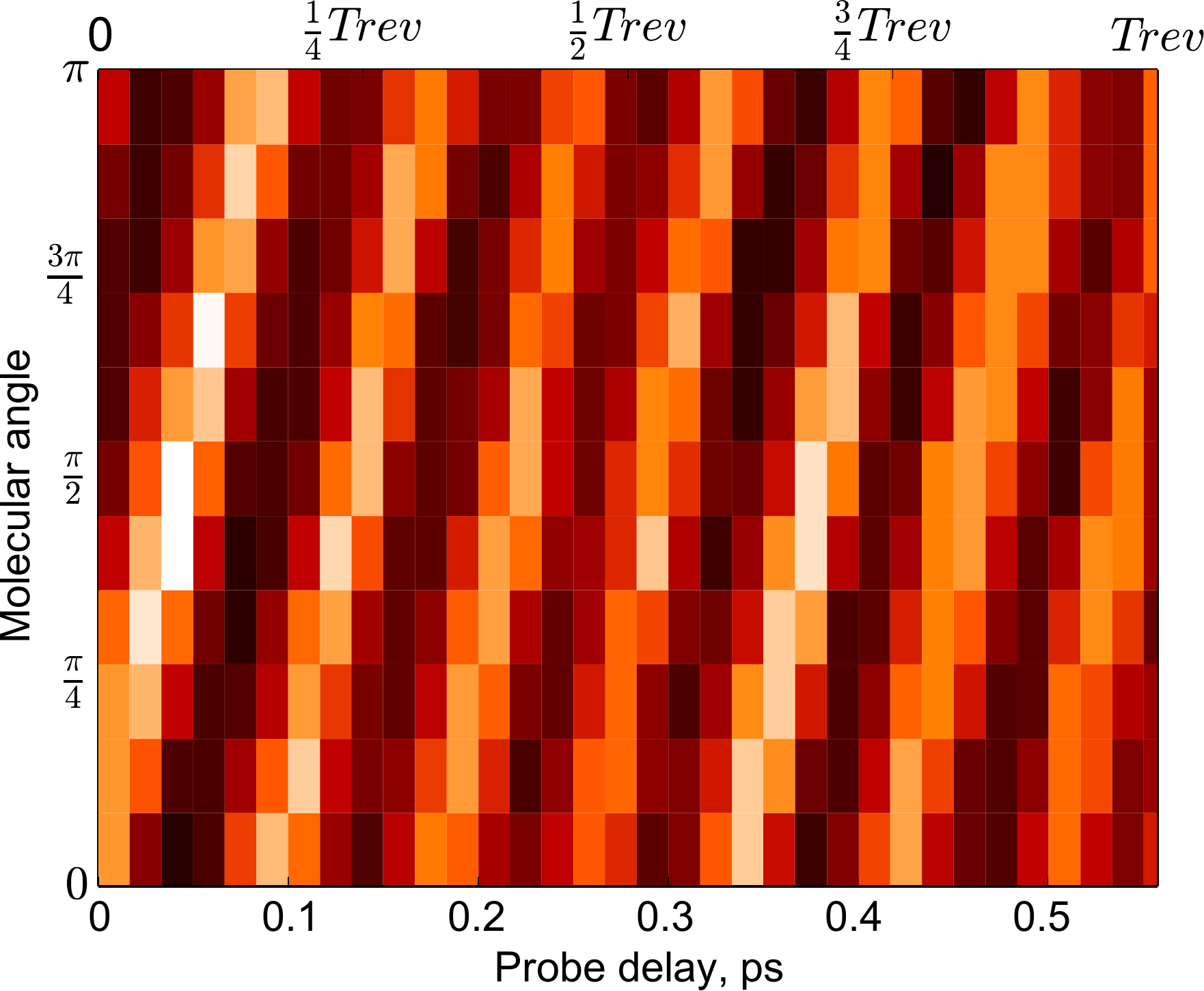}
\caption{\textbf{(a)} Probability density as a function of the molecular angle and the the free propagation time of D$_{2}$ prepared in the equal-weight superposition of $N=2$ and $N=4$ states. The observed nondispersing behavior illustrates the main property of a quantum cogwheel state.}
\label{fig_deuterium}
\end{figure}

To create a truly nondispersing quantum cogwheel state, one needs to excite a narrower wave packet consisting of only two $N$-states. Instead of controlling the adiabaticity of an optical centrifuge, one can use a gas of lighter molecules with a larger energy spacing between the rotational levels. We have repeated the experiment with molecular deuterium, whose moment of inertia is $\sim20$ times lower than that of oxygen. D$_{2}$ has two spin isomers, ortho and paradeuterium, with only even or only odd rotational quantum numbers in their rotation spectrum, respectively. The terminal frequency of the centrifuge was set at half the frequency of the $N=2\rightarrow N=4$ Raman transition. Under these conditions, orthodeuterium is excited to the coherent superposition of equally populated $N=2$ and $N=4$ states, while paradeuterium is transferred into a single state with $N=3$. As the probability density of the latter does not dependent on the polar angle $\vartheta $, it introduces a homogeneous background which does not affect the observed dynamics of orthodeuterium, shown in \figrefs{fig_deuterium}. One can see that during one revival time, the molecule completes exactly $3\frac{1}{2}$ full rotations in agreement with the expected frequency of classical rotation $\Omega_{2} =\frac{1}{2\pi}\frac{\omega_4-\omega_2}{2}=7\frac{B}{h}=\frac{7}{2}\frac{1}{T_{rev}}$. We observe no dispersion in the angular shape of the created wave packet, as anticipated for the time evolution of a cogwheel state.

\section{Conclusion}
In summary, we have experimentally created and observed rotational wave packets which mimic the classical motion of a rotating object, such as a dumbbell or a cross, which preserves its shape over an extended period of time. We have demonstrated that the classicality of the rotational dynamics depends on the distribution width of the molecular angular momentum, with narrower distributions showing closer resemblance to classical behavior. The technique of an optical centrifuge has enabled us to produce very narrow rotational wave packets in different molecular species. In case of deuterium, where the created rotational wave packet consisted of only two rotational states, a truly nondispersing classical-like dynamics of a cogwheel state has been observed. A slightly broader distribution of angular momentum in centrifuged oxygen resulted in richer dynamics, exhibiting gradual transitions between the molecular wave functions of different spatial symmetries.

This work has been supported by the CFI, BCKDF and NSERC, and carried out under the auspices of the Center for Research on Ultra-Cold Systems (CRUCS). We gratefully acknowledge stimulating discussions with E. A. Shapiro.

\end{document}